\documentclass{JHEP3}
\usepackage{graphicx}
\usepackage{amssymb}
\usepackage[mathscr]{eucal}
\usepackage{amsmath}
\usepackage{cite}
\usepackage{afterpage}
\usepackage{slashed}
\usepackage{feynmp}

\newcommand{\gsimm}{\raise.3ex\hbox{$>$\kern-.75em\lower1ex\hbox{$\sim$}}}
\newcommand{\lsimm}{\raise.3ex\hbox{$<$\kern-.75em\lower1ex\hbox{$\sim$}}}

\newcommand{\be}{\begin{equation}}
\newcommand{\ee}{\end{equation}}
\newcommand{\ba}{\begin{eqnarray}}
\newcommand{\ea}{\end{eqnarray}}

\newcommand{\bea}{\begin{eqnarray*}}
\newcommand{\eea}{\end{eqnarray*}}

\title{Atomic Interferometry Test of Dark Energy}

\author{Philippe Brax\\
Institut de Physique Th\'{e}orique, Universit\'e Paris-Saclay, CEA, CNRS, F-91191 Gif/Yvette Cedex, France
 \\ E-mail:
  \email{philippe.brax@cea.fr}}

\author{Anne-Christine Davis\\
  DAMTP, Centre for Mathematical Sciences, University of Cambridge,
  CB3 0WA, UK\\E-mail:
  \email{A.C.Davis@damtp.cam.ac.uk}}

\date{today}
\abstract{Atomic interferometry can be used to probe dark energy models coupled to matter. We consider the constraints coming from recent experimental results on models generalising the inverse power law chameleons such as $f(R)$ gravity in the large curvature regime, the environmentally dependent dilaton and symmetrons. Using the tomographic description of these models, we find that only symmetrons with masses smaller than  the dark energy scale can be efficiently tested. In this regime, the resulting constraints complement the bounds from the E\"otwash experiment and exclude small values of the symmetron self-coupling.  }

\begin{document}
\section{Introduction}
Dark energy \cite{Copeland:2006wr} has proved to be elusive since its discovery some fifteen years ago \cite{Riess:1998cb, Perlmutter:1998np}. The only tangible proof of its existence follows from a host of cosmological observables ranging from the original SNIa supernovae data to the more recent results obtained by the Planck mission \cite{Ade:2015xua} and the observation of baryonic acoustic oscillations \cite{Aubourg:2014yra}. A better understanding of its nature would wish to complement indirect evidence with experimental data in the laboratory \cite{Adelberger:2003zx, Lamoreaux:1996wh,Nesvizhevsky:2003ww,Jain:2013wgs} in a way akin to what has been attempted over the last decades for dark matter. Effects of dark energy on small scale experiments require the presence of a coupling to matter, and as a result the dark energy models with possible experimental tests in the laboratory fall within the class of dark energy/modified gravity theories \cite{Clifton:2011jh}. They have been recently classified according to the type of screening mechanism that shields the dark energy interaction with matter in our local, i.e. solar, environment \cite{Joyce:2014kja}. There are three broad families of such models: the ones subject to the Vainshtein \cite{Vainshtein:1972sx} or K-mouflage mechanisms \cite{Babichev:2009ee}, or the generalised chameleon property \cite{Khoury:2003aq,KhouryWeltman,Damour:1994zq,Pietroni:2005pv,Olive:2007aj,Hinterbichler:2010es,Brax:2010gi,Brax:2012gr}. The latter is the one which will concern us in this paper. In a nutshell, the chameleon screening occurs in regions of space where the Newtonian potential is large enough. This can occur in two typical ways. The first, and this is the original chameleon mechanism, is where  the dark energy field becomes massive enough in the presence of matter\cite{Khoury:2003aq}. The second is the Damour-Polyakov property \cite{Damour:1994zq} whereby the interaction coupling between dark energy and matter becomes very small in dense matter. Both types of models can be mathematically described using a tomographic method \cite{Brax:2011aw} whereby the coupling function and potential can be reconstructed from the sole knowledge of the density dependence of both the mass and the interaction coupling.

Laboratory tests of dark energy have been considered in the last ten years with wide-ranging techniques, see for instance \cite{Rider:2016xaq,Burrage:2016lpu} and a summary of the bounds on chameleons \cite{Burrage:2016bwy} and modified gravity models \cite{Lombriser:2014dua}. Stringent constraints \cite{Mota:2006ed,Mota:2006fz,Brax:2008hh} follow from torsion pendulum experiments such as  E\"otwash \cite{Hoyle:2004cw,Kapner:2006si} where the presence of new forces can be tested \cite{Upadhye:2012qu}. Another promising technique uses the potential deviation from the Casimir interaction \cite{Lamoreaux:1996wh,Decca:2007yb} between two plates \cite{Brax:2007vm,Almasi:2015zpa}. Forthcoming experiments such as CANNEX may potentially exclude all models of the inverse chameleon type \cite{Brax:2010xx}. Finally neutrons can be efficiently used \cite{Nesvizhevsky:2003ww}. First of all, the energy levels of neutrons in the terrestrial gravitational field have been measured  \cite{Nesvizhevsky:2003ww,Jenke:2014yel} and deviation from this pattern would signal the existence of new interactions of the inverse chameleon type \cite{Brax:2011hb,Brax:2013cfa,Schmiedmayer:2015cqa}. Neutron interferometry can also be implemented as new interactions of the chameleon type \cite{Brax:2013cfa} would induce a phase shift and therefore a change in the interferometric patterns \cite{Lemmel:2015kwa}. More recently, atomic interferometry \cite{Burrage:2014oza} has been suggested as a new technique for probing dark energy. Experimental results have already been obtained \cite{Hamilton:2015zga} and constraints on inverse power law chameleons deduced \cite{Elder:2016yxm,Schlogel:2015idt}. In this paper, we will generalise this analysis to all models described by the tomographic method and therefore subject to either the chameleon or the Damour-Polyakov screening mechanisms \cite{Brax:2014zta}. This captures interesting models such as $f(R)$ gravity in the large curvature regime \cite{Hu:2007nk}, the environmentally dependent dilaton \cite{Brax:2010gi} and the symmetron \cite{Hinterbichler:2010es}. We find that the only models which can be efficiently tested by atomic interferometry are the symmetrons with masses falling below the dark energy scale. Symmetrons with mass larger than the present Hubble rate are known to have relevant implications cosmologically, but cannot be tested by this method \cite{Brax:2011aw}. On the other hand, symmetrons with masses of order of the dark energy scale are within reach of the E\"otwash types of experiments \cite{Upadhye:2012rc}. Here we find that atomic interferometry can probe symmetrons with masses a few orders of magnitude below the dark energy scale, typically with a range in vacuum smaller than a few centimeters.

The paper is arranged as follows. In section 2, we recall details of the tomographic method and its link to models such as inverse power law chameleons, $f(R)$ gravity in the large curvature regime, the environmentally dependent dilaton and the symmetron. In section 3, we provide analytical details about scalar fields in a cylinder as suited to analyse current experimental data. In section 4, we apply the tomographic method to atomic interferometry experiments. Finally in section 5, we give details about constraints on models and we focus on symmetrons. Conclusions are in section 6. There are three appendices where we compute the
field profile in the cavity, the force profile and scalar charge and the
E\"otwash bounds using the tomographic method for the symmetron.

\section{Tomographic Models}

\subsection{The tomographic method}
In this paper, we shall focus on
inverse power law chameleons, the only case for which atomic interferometry constraints have been given,  and their generalisations. All these models are scalar-tensor theories described by the Lagrangian
\be
S=\int d^4 x \sqrt{-g}(\frac{R}{16\pi G_N} -\frac{(\partial \phi)^2}{2} -V(\phi))+S_m (\psi, A^2(\phi) g_{\mu\nu})
\label{act}
\ee
where $A(\phi)$ specifies the coupling between matter fields $\psi$ and the scalar $\phi$. The coupling to matter itself  is given by the derivative
\be
\beta (\phi)= m_{\rm Pl} \frac{d \ln A(\phi)}{d \phi}.
\ee
A salient feature of these models  is that the  dynamics are determined by an effective potential which takes into account the presence of the conserved matter density $\rho$ of the environment
\be
V_{\rm eff}(\phi) =V(\phi) +(A(\phi)-1) \rho.
\label{eff}
\ee
All the tomographic models \cite{Brax:2011aw} are obtained when the effective potential acquires a matter dependent minimum $\phi_{}(\rho)$, for instance when $V(\phi)$ decreases and $A(\phi)$ increases. At the minimum of the effective potential,
the mass of the scalar becomes also matter dependent $m(\rho)$.
In this case, all scalar-tensor theories  can  be described
parametrically only from the knowledge of the mass function $m(\rho)$ and the coupling $\beta (\rho)$ at the minimum of the potential \cite{Brax:2012gr,Brax:2011aw}. In the following, we shall use the  simpler description where  the functions $m(\rho)$ and $\beta(\rho)$ become function of the scale factor of the Universe $a$ using the mapping of the matter density
\be
\rho(a)=\frac{\rho_0}{a^3}
\label{dens}
\ee
where $\rho_0= 3 \Omega_{m0} H_0^2 m_{\rm Pl}^2$, $a\le 1$ and $a_0=1$ today. This allows one to describe  models in a simple way. The field value is given by
\be
\frac{\phi (a)-\phi_i}{m_{\rm Pl}}= 9\Omega_{m0} H_0^2 \int_{a_i}^{a} da \frac{\beta (a) }{a^4m^2(a)},
\label{tom}
\ee
where the Hubble rate now is $H_0\sim 10^{-43}$ GeV and the matter fraction is $\Omega_{m0}\sim 0.27$. The choice of the lower bound $a_i$ only shifts the value of the constant $\phi_i$.
The mass function is identified as the second derivative
\be
m^2 (a)= \frac{d^2 V_{\rm eff}}{d\phi^2}\vert_{\phi=\phi (\rho(a))}
\ee
and the coupling to matter is given by
\be
\beta (a)= m_{\rm Pl} \frac{d\ln A}{d\phi}\vert_{\phi=\phi(\rho(a))}.
\ee
The potential can also be reconstructed and  is given by
\be
V(a)-V_i= -27\Omega_{m0}^2 H_0^4 \int_{a_i}^{a} da \frac{\beta^2 (a) m_{\rm Pl}^2}{a^7m^2(a)}.
\label{tom1}
\ee
where $V_i$ is a constant.
Eliminating $a$ amongst these expressions allows one  to obtain $V(\phi)$ and $A(\phi)$ implicitly from $m(a)$ and $\beta (a)$. Whilst these equations have
been written in terms of cosmological parameters, $\Omega_{m0}$, $H_0$, this is
for calculational convenience and does not imply the parametrisation is only
valid cosmologically. Indeed, we have used this previously when applying laboratory constraints to other modified gravity models \cite{Brax:2014zta}. The
parametrisation very efficiently encompasses the range of models being tested and
the typical mass parameter of the dark energy scale.

In the following subsections, we will give details about the tomographic method for many popular models ranging from the inverse power chameleon to the symmetron. In a nutshell, we will give the $(m(a),\beta (a))$ parameterisation and use (\ref{tom}) and (\ref{tom1}) to calculate both $\phi(a)$ and $V(a)$. Eliminating $a$ will give the dependence $V(\phi)$. We can also infer $\beta(\phi)$ and by integration $A(\phi)$.
In fact the $(m(a),\beta(a))$ parameterisation is all we will need to compare to atomic interferometry data. In particular, the dependence $\phi(a)$ allows us to determine the variation of the field $\phi(\rho)$ as the function of the matter density at the minimum of the effective potential (\ref{eff}). Indeed, all one needs to do is to use the mapping (\ref{dens}) to get the $\rho$ dependence. Similarly, one can use the tomographic mapping in another way. For instance, the value of the field $\phi_c$ at the centre of a cavity is not directly related to the density in the cavity. It is a function of the size of the cavity (see (\ref{res}) and (\ref{res1}) in section 3). Nevertheless we can associate a scale factor $a_c$ such that $\phi_c=\phi(a_c)$. The values of $a_c$ associated to the field in the centre of a cavity will be given below for each model. We will make all this explicit for the inverse power law chameleons in the next subsection and give fewer details for the other models.

\subsection{Inverse power law chameleons}

The only models which have been used so far when analysing atomic interferometry results \cite{Burrage:2014oza} are the
chameleons with an inverse power law  potential of the type
\be
V(\phi)=\Lambda^4 +\frac{\Lambda^{4+n}}{\phi^n} + \dots
\ee
with $n>0$, $\Lambda\sim 10^{-3}$ eV is the cosmological vacuum energy now, and the coupling function is such that $\beta$ is constant, i.e.
\be
A(\phi)=\exp(\frac{\beta\phi}{m_{\rm Pl}}),
\ee
which implies that
\be
\beta(a)=\beta.
\ee
It is easy to see that using the mass dependence on the scale factor
\be
m(a)= m_0 a^{-r}
\ee
in (\ref{tom}) leads to a power law for $\phi(a)$
\be
\frac{\phi(a)}{m_{\rm Pl}}=\frac{9\Omega_{m0} H_0^2}{m_0^2} \frac{\beta}{2r-3} a^{2r-3}
\ee
and for $V(a)$ in (\ref{tom1})
\be
V(a)= V_0 -\frac{27\Omega_{m0}^2 H_0^4}{m_{\rm Pl}^2 m_0^2}\frac{\beta^2}{2r-6}a^{2r-6}.
\ee
for a constant $V_0$. Eliminating $a$ between these two expressions, we retrieve the inverse power law model where $r= \frac{3(n+2)}{2(n+1)}$ and
the mass scale $m_0$ is determined by
\be
m_0^{2(n+1)}=\frac{(n+1)^{n+1}}{3n}\frac{(3\beta \Omega_{m0}H_0^2 m_{\rm Pl})^{n+2}}{ \Lambda^{4+n} }.
\ee
This implies that inverse chameleon models have a cosmological interaction range $1/m_0$  much shorter than the size of the observable Universe for $\beta_0\gtrsim 1$. In the following, we will generalise this simple
parameterisation to more complex models. The same method can be applied to all the models presented below.

\subsection{Large curvature f(R)}
Chameleon models involve a scalar field. Surprisingly, some models of modified gravity which do not seem to involve a scalar field can in fact be mapped to scalar-tensor theories.
This is the case of
a large class of interesting models such as  the large curvature $f(R)$ models with the action \cite{Hu:2007nk}
\be
S=\int d^4x \sqrt{-g} \frac{f(R)}{16\pi G_N}
\ee
involving  the function $f(R)$ which is expanded in the large curvature regime
\be
f(R)=\Lambda_0 + R -\frac{f_{R_0}}{n} \frac{R_0^{n+1}}{R^n}.
\ee
Here $\Lambda_0$ is the cosmological constant term  leading  to the late time acceleration of the Universe and $R_0$ is the present day curvature.  These models can be described using the constant  $\beta(a)=1/\sqrt{6}$ and the corresponding mass function
\be
m(a)= m_0 (\frac{4\Omega_{\Lambda 0}+ \Omega_{m0} a^{-3}}{4\Omega_{\Lambda 0}+ \Omega_{m0}})^{(n+2)/2}
\ee
where the mass on large cosmological scale is given by
\be
m_0= H_0 \sqrt{\frac{4\Omega_{\Lambda 0}+ \Omega_{m0} }{(n+1) f_{R_0}}},
\ee
and $\Omega_{\Lambda 0} \approx 0.73$ is the dark energy fraction now \cite{Brax:2012gr}. When $a\ll 1$ corresponding to dense environments, the mass dependence on $a$ is a power law
\be
m(a) \sim m_0 a^{-r}
\ee
where $r=\frac{3(n+2)}{2}$.
\subsection{ Generalised power law models}

The inverse power law chameleons, and f(R) models in the large curvature limit  are  described by power law functions of the scale factor
\be
m(a)= m_0 a^{-r}, \ \beta(a)=\beta_0 a^{-s}
\ee
for different choices of $r$ and $s$ \cite{Brax:2014zta}. In fact, all these models are equivalently defined by   power law  potentials
\be
V(\phi) = V_0 +\epsilon \Lambda_p^{4-p}\phi^p
\ee
 where $V_0$ is a constant, and the exponent is given by
\be
p=\frac{2r-6-2s}{2r-3-s}
\ee
as long as $(2r-3-s)>0$. The sign $\epsilon=\pm 1$ is positive when $p<0$ and vice versa. As for inverse power law chameleon models, it is convenient to  introduce the effective scale
\be
\Lambda_p^{4-p}=\frac{27}{\vert 2r-6-2s\vert } \frac{\Omega_{m0}^2 \beta_0^2 H_0^4 m_{\rm Pl}^2}{m_0^2} (\frac{2r-3-s}{9} \frac{m_0^2}{ \Omega_{m0} \beta_0 H_0^2 m_{\rm pl}})^p
\ee
which is a function of both $m_0$ and $\beta_0$. Using these ingredients, we find that the coupling function is given by
 \be
A(\phi)= 1+ \frac{\beta_0}{m_{\rm Pl}} \frac{\phi^l}{M^{l-1}}
\ee
where the power $l$ is simply
\be
l=\frac{2r-3-2s}{2r-3-s}
\ee
 and the coupling scale is
\be
M^{1-l}= \frac{\Omega_{m0}}{l} (\frac{9}{2r-3-s}\frac{\Omega_{m0}\beta_0 H_0^2 m_{\rm Pl}}{m_0^2})^{\frac{s}{2r-3-s}}.
\ee
This field theoretic parameterisation is only given here as an illustration since the $(m(a),\beta (a))$ description is far easier to use.

\subsection{Dilaton}
The chameleon and $f(R)$   models are screened by the chameleon mechanism. We will now give examples where the Damour-Polyakov mechanism is at play \cite{Damour:1994zq}.
This is the case of the environmentally dependent dilaton \cite{Brax:2010gi} which is inspired by string theory in the large string coupling limit and has
an exponentially runaway potential
\be
V(\phi)=V_0 e^{-\frac{\phi}{m_{\rm Pl}}}
\ee
where $V_0$ is determined to generate the acceleration of the Universe now and the coupling function is quadratic
\be
A(\phi)=\frac{A_2}{2m_{\rm Pl}^2} (\phi-\phi_\star)^2.
\ee
These models can be described using the coupling function
\be
\beta(a)= \beta_0 a^3
\ee
where $\beta_0$ is related to $V_0$ and is determined by requiring that $\phi$ plays the role of late time dark energy which sets $\beta_0=\frac{\Omega_{\Lambda 0}}{\Omega_{m0}}\sim 2.7$, and the mass function which reads
\be
m^2(a)= 3 A_2 \frac{H_0^2}{a^3}
\ee
and is proportional to the Hubble rate with the mass on cosmological scales now given by $m_0=\sqrt{3 A_2} H_0$.

\subsection{Symmetrons}

The symmetron \cite{Hinterbichler:2010es} uses a similar type of coupling function as the dilaton with a quartic potential with a non-vanishing minimum
\be
V(\phi)= V_0 + \frac{\lambda}{4} \phi^4 -\frac{\mu^2}{2} \phi^2.
\ee
The coupling function is chosen to be
\be
A(\phi)= 1 + \frac{ \phi^2}{M_\star^2}
\ee
where the transition from the minimum of the effective potential  at the origin to a non-zero value happens at $a=a_\star$ where
\be
\rho_\star= M_\star^2 \mu^2
\ee
corresponding to
\be
\rho_\star= \frac{\rho_0}{a_\star^3}.
\ee
The symmetrons are defined by the three parameters $(\lambda,\mu,M_\star)$. In the following,
it will be  convenient to change these parameters and to define
\be
m_\star= \sqrt 2 \mu,\ \phi_\star= \frac{2\beta_\star \rho_\star}{m_\star^2 m_{\rm Pl}},\
\ee
where
\be
\lambda= \frac{\mu^2}{\phi_\star^2}.
\ee
 The
symmetron  model can be reconstructed using
\be
m(a)=m_\star \sqrt{ 1-(\frac{a_\star}{a})^3}
\ee
and
\be
\beta(a)=\beta_\star \sqrt{1-(\frac{a_\star}{a})^3}
\ee
for $a>a_\star$ and $\beta (a)=0$ for $a<a_\star$. In a  dense environment, the field is at the origin while in a sparser one with $a>a_\star$ we have
\be
\phi= \phi_\star \sqrt{1-(\frac{a_\star}{a})^3}.
\ee
When the field is at the origin, the mass squared becomes $m^2= (\frac{\rho}{\rho_\star}-1) \mu^2$.
In the tomographic parameterisation, the three parameters of the symmetron models are now $(m_\star,\phi_\star,a_\star)$.
For cosmological applications, it is customary to consider that $\mu \gtrsim 10^3 H_0$ whilst previous laboratory searches have focused on $\mu \sim \Lambda$. Here we will see that atomic interferometry is only sensitive to values of $\mu$ smaller than $\Lambda$.

\section{Scalar Field in a Cylinder}

\subsection{The scalar field profile}

Atomic interferometry experiments constrain the extra acceleration that an atom of (typically) caesium  may experience inside the interferometer when interacting with a source (typically an aluminium ball). The  experiment takes place in a cylindrical cavity and the scalar acceleration depends on the value taken by the scalar field at the centre of the cavity. In this section, we will analyse this situation using known results \cite{Brax:2007hi}.
We consider an infinite cylinder of radius $R_c$ filled with a gas of density $\rho_{\rm in}$ surrounded by a bore of high density $\rho_\infty$. The minimum values of the field in these environments
are respectively $\phi_{\rm in}$ and $\phi_\infty$. Inside the cylinder, the field has a value close to $\phi_c$ which is reached in the centre of the cylinder. The mass of the scalar in the bore is assumed to satisfy
$m_\infty R_c \gg 1$. This leads to the self-consistency equation for the field $\phi_c$ \cite{Brax:2007hi}
\be
\phi_\infty- \phi_c= \frac{\frac{dV_{\rm eff}}{d\phi}\vert_{\phi_c}}{m^2_c} (J_0(im_cR_c)-1)
\label{e1}
\ee
where $J_0$ is the Bessel function of zeroth order.
Using the tomographic method which associates to $\phi_c$ a value of the scale factor $a_c$, we have the expression
\be
\frac{dV_{\rm eff}}{d\phi}\vert_{\phi_c}=(\rho_0 -\rho_c) \frac{\beta_c }{m_{\rm Pl}}
\label{e2}
\ee
where $\rho_c= \frac{\rho_0}{a_c^3}$ by definition of  $\phi_c=\phi(a_c)$. We can simplify the analysis using that $\rho_{\rm in}\ll \rho_c$  and $\phi_{c}\gg \phi_\infty$,  and we find that
\be
J_0(im_cR_c)=1 + \frac{\phi_c m_c^2 m_{\rm Pl}}{\beta_c \rho_c}
\label{res}
\ee
This determines the values of $m_c$ as a function of $R_c$ and leads to a resonance condition.

\subsection{The resonance condition}

We are going to analyse (\ref{res}) for the tomographic models presented in section 2.
Let us first consider the case of generalised power law models. Using (\ref{tom})
\be
\phi_c \sim \frac{9\Omega_{m0} \beta_0 H_0^2}{m_0^2} \frac{a_c^{2r-3-s}}{2r-3-s}
\ee
we find that the field inside the cylinder must satisfy the resonance condition
\be
J_0(im_cR_c)=1+ \frac{3}{2r-3-s}
\ee
and as result we expect that
\be
m_c R_c = {\xi}
\label{res}
\ee
where  the parameter $\xi$ is determined by
\be
J_0(i\xi)= 1+ \frac{3}{2r-3-s}.
\ee
For inverse power law chameleons we find that
\be
J_0(i\xi)=n+2
\ee
while for large curvature $f(R)$ it becomes
\be
J_0(i\xi)= \frac{n+2}{n+1}
\ee
and for dilatons
\be
J_0(i\xi)=2.
\ee
In all these cases we have that $m_c R_c={\cal O}(1)$ \cite{KhouryWeltman, Elder:2016yxm,Hamilton:2015zga} implying that the range of the scalar force is of the order of the size of the cavity. This guarantees that $\phi_{c}\gg \phi_\infty$ as the vacuum range of the scalar interaction is much larger than the cavity.

The symmetron behave significantly differently and we find that the resonance condition reads
\be
J_0(im_cR_c)= \frac{1+ \frac{m_c^2}{m_\star^2}}{1- \frac{m_c^2}{m_\star^2}}.
\ee
This equation admits a solution when $m_\star R_c\lesssim 1$ corresponding to a force whose cosmological range is larger than the size of the cavity
\be
m_cR_c= 4\sqrt{\frac{8}{m_\star^2 R_c^2} -1}\sim \frac{8\sqrt 2}{{m_\star R_c}}.
\ee
This implies that the range of the force inside the cavity is smaller than the size of the cavity.
In particular we have that
\be
\sqrt{1-\frac{a_\star^3}{a_c^3}}\sim \frac{8\sqrt 2}{(m_\star R_c)^{2}}\gg 1
\ee
which is obviously a contradiction. So when $m_\star R_c \lesssim 1$, we find the field inside the cavity vanishes like in the bore
\be
\phi_c=0
\ee
which is a solution of (\ref{e1}) and (\ref{e2}) when $\beta_c=0$
This phenomenon is similar to the one already obtained in the 1d case between infinite plates \cite{Upadhye:2012rc,Brax:2014zta}.

When $m_\star R_c \gg 1$, i.e. when the cosmological range is smaller than  the cavity
 the solution is such that
\be
m_c=m_\star(1 - \frac{1}{2}\sqrt{\frac{\pi}{2} m_\star R_c} e^{-m_\star R_c})
\label{res1}
\ee
which is exponentially close to $m_\star$. This implies that the range of the symmetron inside the cavity is essentially given by $1/m_\star$, i.e. the cosmological one which is smaller than the size of the cavity.

In both cases, the relations (\ref{res}) and (\ref{res1}) correspond to a scale factor $a_c$ from which one can calculate the value of the field $\phi_c$ inside the cavity as
\be
\phi_c\equiv \phi(a_c)
\ee
which is obtained using (\ref{tom}). Hence the tomographic method allows us to calculate the value of the field $\phi_c$ for all tomographic models.

\section{Atomic Interferometry}

\subsection{Experimental constraints}

The atomic interferometry experiments constrain the anomalous acceleration of an atom in the terrestrial gravitational field in the presence of an external ball of matter. They provide relevant tests of screening mechanisms. Indeed the external ball induces an extra acceleration compared to the Newtonian one with \cite{Hamilton:2015zga,Elder:2016yxm}
\be
a_S \lesssim 5.5 \mu m/s^2
\ee
at a distance $d=R_S+ d_S$ where $R_S=0.95$ cm is the radius of the ball and $d_S=0.88$ cm is the distance to the interferometer. The whole apparatus is embedded inside a cavity of radius $R_c=6.1$ cm.
The acceleration due to the scalar is given by (see the appendix for a general discussion)
\be
a_B= 2 Q_S Q_B\frac{G_N m_S}{d^2}= 2 Q_S Q_B \frac{\Phi_N R_S}{d^2}
\ee
where $m_S$ is the mass of the source and $\Phi_N$ is the Newtonian potential at the surface of the ball. A good approximation for the scalar charges $Q_S$ and $Q_B$ is obtained by considering that in the screened case the value of the scalar is constant inside the object, leading to
\be
Q_A= \frac{\vert \phi_A - \phi_c \vert}{2m_{\rm Pl} \Phi_N}
\ee
when the object $A$ is screened, i.e. when
\be
Q_A \le \beta_c
\ee
where $\beta_c$ is the coupling in the vacuum of the cavity and $\phi_c$ the field there. We have denoted by $\phi_A$ the value of the field inside the screened object.
If the object is not screened then
\be
Q_A=\beta_c.
\ee
The value of the field inside the cavity is such that
\be
m_c=\frac{\xi}{R_c}
\ee
where $\xi$ is a fudge factor of order one which must be fitted to more precise numerical simulations, see the previous section for a theoretical discussion and below where $\xi$ is fitted to the actual experiment setup in the chameleon case \cite{Elder:2016yxm}. This determines a scale factor $a_c$ characteristic of the cavity. For symmetrons we have $m_c\sim m_\star$.

These constrains are only valid when the Yukawa suppression factor in the exact expression for the acceleration
\be
a_B= 2 Q_S Q_B \frac{\Phi_N R_S}{d^2} e^{-m_c d_S}
\ee
can be safely put to one, i.e when
\be
m_c d_S \ll 1
\ee
which occurs when the experiment is designed such that
\be
R_c\gg \xi d_S.
\ee
The current experimental results are obtained for $R_c/d_S\sim 6$ which requires a rather small value of $\xi$.
\subsection{Screening of the nucleus}

The comparison with the experimental constraint requires one to know whether the atom and in particular its nucleus is screened. The nucleus is screened when
\be
\vert \phi_c -\phi_B \vert \le 2 m_{\rm Pl} \beta_c \Phi_B
\ee
where $\Phi_B$ is the Newtonian potential at its surface. For a caesium atom this is around $\Phi_B \sim 10^{-38}$.
This criterion can be rewritten as
\be
\frac{9\Omega_{m0} H_0^2}{2\Phi_B}\int_{a_B}^{a_c} da \frac{\beta(a)}{a^4 m^2(a)}\le \beta_c
\ee
where $a_B$ corresponds to the nuclear density, i.e. $a_B \ll a_c$ and
\be
\rho_B=\frac{3\Omega_{m0} m_{\rm Pl}^2 H_0^2}{a_B^3}.
\ee
For all power law models, this constraint reads
\be
\frac{9\Omega_{m0} H_0^2}{2\Phi_B m_0^2}  \frac{a_c^{2r-3}}{2r-3-s}\le 1
\ee
whilst for the symmetron, assuming that $a_c>a_\star$ and $a_B<a_\star$
\be
\frac{\phi_\star}{2m_{\rm Pl} \Phi_B} \le \beta_\star.
\ee
i.e.
\be
M_\star^2\equiv \frac{\rho_\star}{\mu^2} \le 2 m_{\rm Pl}^2 \Phi_B.
\ee
For dilatons we find that
\be
A_2\ge \frac{\Omega_{m0}}{2\Phi_B}
\ee
implying that cosmologically interesting dilatons where $A_2 \sim 10^6$ are such that the nucleus is always unscreened.
In the case of large curvature $f(R)$ models we find that the nucleus is screened for
\be
f_{R_0}\le (\frac{2A^2 (n+1)\Phi_B}{3\Omega_{m0}})^{n+2} (AH_0 R_c)^{-2(n+1)}
\ee
where
\be
A= \sqrt{\frac{4\Omega_{\Lambda0}+ \Omega_{m0}}{n+1}}.
\ee
Numerically  for all models with $n\ge 1$, the nucleus is screened when $f_{R_0}\lesssim 1$. Similarly chameleons are screened inside the nucleus when
\be
\beta_0 \ge \frac{3\Omega_{m0}(n+1)}{2\Phi_B}\frac{H_0^2}{M_0^2} (\frac{M_0 R_c}{\xi})^{2/(n+2)}
\ee
with
\be
M_0= (\frac{(n+1)^{n+1}}{3n} \frac{(3\Omega_{m0}H_0^2 m_{\rm Pl})^{n+2}}{\Lambda^{n+4}})^{1/2(n+1)}
\ee
which is larger than the current bounds on $\beta_0$. Thus we will consider the nucleus to be unscreened.
Finally for generalised power law models
\be
\frac{9\Omega_{m0}}{2\Phi_B}\frac{H_0^2}{m_0^2} \frac{(m_0R_c \xi^{-1})^{(2r-3)/r}}{2r-s-3} \le 1
\ee
for screening. The nucleus is always screened for large enough values of $r$.

\section{Constraints on Models}

\subsection{Model independent constraint}
Let us assume that the atoms are not screened and the source is screened. The scalar charge of the source becomes
\be
Q_S= \frac{9\Omega_{m0} H_0^2}{2\Phi_N}\int_{a_S}^{a_c} da \frac{\beta(a)}{a^4 m^2(a)}
\ee
where the density of the source can be parameterised as
\be
\rho_S= \frac{3\Omega_{m0} m_{\rm Pl}^2 H_0^2}{a_S^3}.
\ee
In this case the interferometry constraint reads
\be
{9\Omega_{m0} H_0^2}\int_{a_S}^{a_c} da \frac{\beta(a)}{a^4 m^2(a)}\le C\beta_c^{-1}
\ee
where
\be
C=5.5 \frac{d_S^2}{R_S} \mu {\rm m/s^2} \sim 2. 10^{-24}
\ee
is a pure number.
When the nucleus is screened, the constraint changes and becomes
\be
{9\Omega_{m0} H_0^2}\int_{a_S}^{a_c} da \frac{\beta(a)}{a^4 m^2(a)}\le \sqrt{2C\Phi_B}.
\ee

\subsection{Power law models}

For power law models we have
\be
a_c= (\frac{m_0R_c}{\xi})^{1/r}
\ee
and as long as $a_S\ll a_c$, we find the constraint
\be
2 \Phi_N Q_S \beta_c= \frac{9\Omega_{m0} \beta_0^2}{2r-3-s} (\frac{H_0}{m_0})^2 a_c^{2r-3-2s} \le C
\ee
where $2r-3-s>0$.
This is the case when the atoms are not screened.
When the nucleus is screened, the constraint becomes slightly different
\be
\frac{9\Omega_{m0} \beta_0}{2r-3-s} (\frac{H_0}{m_0})^2 a_c^{2r-3-s}\le  \sqrt{2C\Phi_B}
\ee
For different power law models with $m_0=10^3 H_0$ and $s=-1,-0.5,0,0.5,1$ we have plotted in figure 1 the upper bound on $\beta_0$ as a function of $r$ assuming that $\xi=1$. We can see that models with $\beta_0={\cal O}(1)$ are excluded when $s>0$ and models with $s\le 0$ are excluded when $r$ is large enough.
\begin{figure}
\centering
\includegraphics[width=0.50\linewidth]{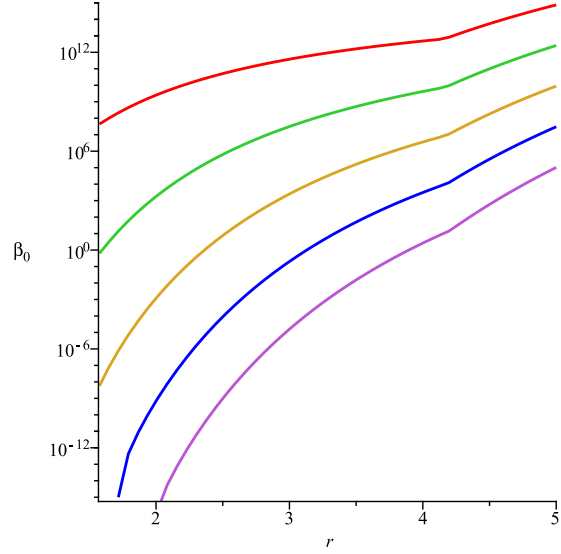}
\caption{The upper bound on the coupling $\beta_0$ for power law models as a function of the index $r$ for $s=-1,-0.5,0,0.5,1$ from top to bottom. For large $r$ and large $\beta_0$, the nucleus of the test atom is screeneed. Models with $s>0$ and $\beta_0={\cal O}(1)$ are excluded whilst models with $s\le 0$ are excluded at large $r$. }
\end{figure}
We can also specialise to other well-known  models.
\subsubsection{Chameleons}
In this case, the mass $m_0$ is related to $\Lambda$, $n$ and $\beta_0$ and we obtain the bound
\be
\beta_0\le \frac{M_0^2}{H_0^2} (\frac{M_0 R_c}{\xi})^{-\frac{2}{n+2}} \frac{Ce^{d_S \xi/R_c}}{3\Omega_{m0}(n+1)(1+\xi \frac{R_S}{R_c})}
\ee
where $\xi$ has been numerically fitted according to \cite{Elder:2016yxm}
\be
\xi= \xi_J^{(n+2)/2}
\ee
where $0.55\lesssim \xi_J\lesssim 0.65$ has been used . The effect of changing $\xi_J$ is shown in figure 2. We have also reinstated the Yukawa suppression factor as it is not completely irrelevant with the current experimental setup.
\begin{figure}
\centering
\includegraphics[width=0.50\linewidth]{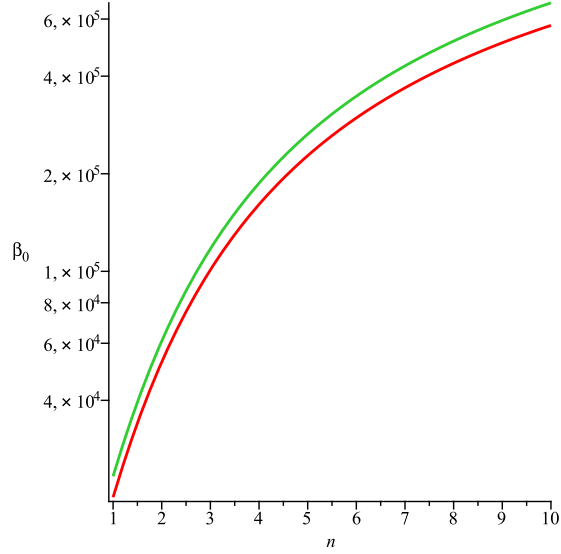}
\caption{The upper bound on the coupling $\beta_0$ for chameleons as a function of the index $n$ for the two extreme values of $\xi_J=0.55$ and $\xi_J=0.65$. }
\end{figure}
This upper bound on $\beta_0$ is displayed in figure 2 and shows that $\beta_0\lesssim 10^5$ as already obtained in \cite{Elder:2016yxm}. The agreement between our analytical results and the numerical analysis of \cite{Elder:2016yxm} is good.

\subsubsection{Large curvature f(R)}

In this case we obtain a bound on $f_{R_0}$ for different values of $n$ using the fact that the nucleus is always screened,
We find that
\be
f_{R_0} \le ( \frac{2 (n+1)A^2 \sqrt{2C\Phi_B}}{9\Omega_{m0}\beta_0} )^{n+2} ( A H_0^2 R_c)^{-2(n+1)}
\ee
which gives $f_{R_0}\le 10^{20} $ for $n=1$ and even looser bounds for larger $n$. Hence the atomic interferometry bound is not effective for $f(R)$ models.

\subsection{Environmentally dependent dilaton}

In this case, the bound implies that $A_2$ is bounded by
\be
A_2 \le \frac{1}{9\Omega_{m0} \beta_0^2} \frac{C}{(H_0 R_c)^4}
\ee
 where $\beta_0= \frac{\Omega_{\Lambda 0}}{\Omega_{m0}}$. This is also not effective as this means $A_2 \lesssim 10^{85}$, which is much larger than the cosmological value.
\begin{figure}
\centering
\includegraphics[width=0.50\linewidth]{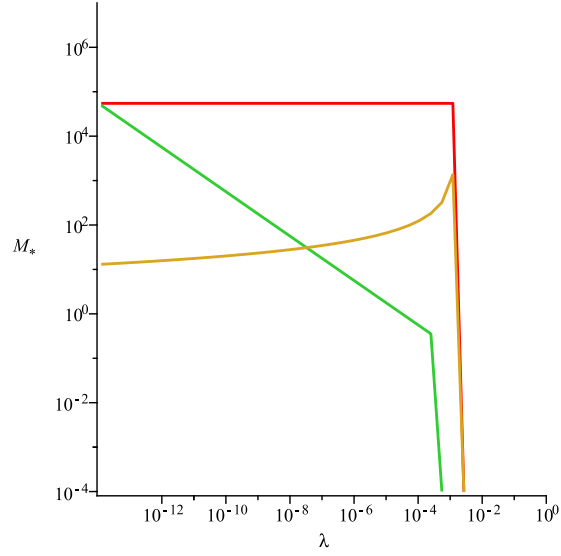}
\caption{The parameter space of symmetrons with $\mu=0.05\Lambda$ as a function of $(\lambda, M_\star)$. The portion of parameter space between the horizontal brown and red curves (middle and top) is excluded by the E\"otwash experiment. Notice that the excluded region is a good approximation to the corresponding exclusion plot obtained using numerical simulations in \cite{Upadhye:2012rc}. The interferometry experiment excludes all the region to the left of the blue curve (leftmost one). Regions with very low values of $M_\star$ below the ones represented here are such that the vacuum is always in the symmetric phase where no effect can be measured.}
\label{syy}
\end{figure}

\subsection{Symmetron}

In the following, we will assume that $m_\star R_c\gtrsim 1$ and take $\mu=\alpha \Lambda$ which satisfies this criterion for $\alpha \ge 10^{-3}$. Typically we will take $\alpha=0.05$ which corresponds to a range
$m_\star^{-1}\sim 1.1 {\rm mm}$. This is smaller than the size of the cavity in the atomic interferometry case, which implies that the symmetron does not vanish in the cavity. This is a valid result provided the density inside the vacuum chamber is low enough, i.e. $\rho_{\rm in}\le \rho_\star$ which implies that
\be
M_\star \ge (\frac{\rho_{\rm in}}{\mu^2})^{1/2}
\ee
where we have introduced
\be
M_\star\equiv \frac{\rho_\star}{\mu^2}.
\ee
which corresponds to $M_\star \ge 1.4 \ 10^{-4} $ GeV for $\mu\sim 1.2\ 10^{-4} {\rm eV}$ for a density of $\rho_{\rm in}=6.6\ 10^{-17}{\rm g/cm^3}$.
For such value of $\mu$, the E\"otwash constraint applies \cite{Upadhye:2012rc} as the range is well below 6 mm, as discussed in Appendix C. These bounds are reproduced in figure (\ref{syy}).
When the nucleus is not screened we find a bound on the parameters
\be
\frac{\beta_\star \phi_\star}{m_{\rm Pl}} \le \frac{C e^{m_\star d_S}}{1+m_\star R_c}
\ee
where we have reinstated the Yukawa suppression as it plays a crucial role here.
Using the link between the tomographic parameters $(m_\star,\phi_\star, a_\star)$ and the symmetron Lagrangian parameters $(\lambda,\mu,M_\star)$
\be
m_\star =\sqrt 2 \mu, \phi_\star= \frac{\mu}{\sqrt\lambda}
\ee
and
\be
\beta_\star= \frac{\mu m_{\rm Pl}}{ \sqrt \lambda M^2_\star}
\ee
this leads to a bound on the coupling to matter
\be
M_\star^2 \ge \frac{\mu^2(1+m_\star R_c)}{\lambda C} e^{-{m_\star d_S}}.
\ee
This is only valid when the atoms are not screened. When they are,  the bound becomes
\be
\frac{\phi_\star}{m_{\rm Pl}} \le \frac{\sqrt{2C\Phi_B}}{\sqrt{1+m_\star R_c}} e^{m_\star d_S/2}
\ee
implying that
\be
\lambda \ge \frac{\mu^2(1+m_\star R_c)}{m_{\rm Pl}^2 }\frac{1}{2\Phi_B  C} e^{-m_\star d_S}
\ee
where $\Phi_B \sim 10^{-38}$ for a caesium atom. The Yukawa suppression has an important impact on this lower bound. In particular for large values of $\mu$ such as $\mu=\Lambda$ where $\mu d_S\gg 1$, the lower bound is essentially irrelevant. Moreover the mass $\mu$ cannot be pushed to very low value as the cosmological range becomes then bigger than the size of the cavity and therefore no force is exerted on the atoms. This implies that for a given size of the cavity and a given distance between the source and the interferometer, a limited range of masses $\mu$ can be tested by atomic interferometry.
We have plotted in figure \ref{syy} the parameter space of symmetrons with $\mu=0.05\Lambda$ in the $(\lambda, M_\star)$ plane. The interferometry experiment excludes regions of very small $\lambda$.
Notice that for very small $a_\star\lesssim 10^{-7}$, the nucleus is not screened whilst for more interesting cosmological values $a_\star\gtrsim 10^{-7}$, the nucleus is screened.
In conclusion, the symmetrons with very small couplings are not excluded by the E\"otwash experiment whereas they are excluded by atomic interferometry, and thus the two different types of experiments are complementary.

\section{Conclusion}

We have studied how dark energy models coupled to matter subject to the chameleon and Damour-Polyakov screening mechanisms can be tested by atomic interferometry experiments. We have used the tomographic description
of these models. Apart from inverse power law chameleons whose coupling to matter must be less than $10^5$, we find that symmetrons with masses in the sub meV region, corresponding to ranges shorter than a few centimeters can be adequately constrained in  a portion  of their parameter space left untouched by torsion pendulum experiments such as E\"otwash. In particular we find that the symmetrom self coupling must be bounded from below and therefore cannot be arbitrarily small.
Future experiments with better sensitivities will certainly lead to improvements on the bounds presented here. This will also help in constraining and maybe even excluding certain chameleon or symmetron models. In the future dark energy coupled to matter may even be eventually detected by such experiments. Having such tests of dark energy in the laboratory, independently of any cosmological signature,  is certainly a necessity in order to understand better the nature of the dark interactions of the Universe.

\section{Acknowledgements}
We would like to thank C. Burrage and H. M\"uller for discussions and suggestions.
This work is supported in part by the EU Horizon 2020 research and innovation programme under the Marie-Sklodowska grant No. 690575 (PB) and the  STFC UK under grants ST/L000385/1 and ST/L000636/1 (ACD) .
Upon completion of this work, we became aware of \cite{burrage} where similar results on the symmetron case were derived simultaneously.
\appendix

\section{The field profile}

We consider the profile of the scalar field inside and outside a dense object of size $R$ when the scalar  mass inside $m_{\rm in}$ satisfies $m_{\rm in}R \gg 1$.
The scalar field is then given by
\be
\phi= \phi_{\rm in}+ A \frac{\sinh m_{\rm in}r}{m_{\rm in}r}
\ee
inside for $r\le R$ and
\be
\phi= \phi_{\rm out} + B \frac{e^{-m_{\rm out}(r-R)}}{r}
\ee
for $r\ge R$. Imposing continuity of the field and its derivative at the boundary leads to
\be
B=-\frac{A}{m_{\rm in}} \frac{m_{\rm in} R \cosh m_{\rm in} R- \sinh m_{\rm in} R}{1+m_{\rm out}R}
\ee
and
\be
A=(1+m_{\rm out }R)\frac{\phi_{\rm in}-\phi_{\rm out}}{ \cosh m_{\rm in} R + {m_{\rm out}R} \frac{\sin h m_{\rm in} R}{m_{\rm in}R}}
\ee
which can be approximated when $m_{\rm in}R \gg 1$ as
\be
A=(1+m_{\rm out }R)\frac{\phi_{\rm out}-\phi_{\rm in}}{\cosh m_{\rm in} R}
\ee
and
\be
B= -  R(\phi_{\rm out} -\phi_{\rm in}).
\ee
When $m_{\rm out} R\ll 1$ we find that
\be
\phi= \phi_{\rm out} + (\phi_{\rm in}-\phi_{\rm out}) \frac{R}{r}
\ee
outside. In the general case we have
\be
\phi= \phi_{\rm out} + (\phi_{\rm in}-\phi_{\rm out}) \frac{R}{r}e^{-m_{\rm out}(r-R)}.
\ee
In particular, the gradient of $\phi$ is given by
\be
\partial_r \phi=-({1+m_{\rm out}r})(\phi_{\rm in}-\phi_{\rm out}) \frac{R}{r^2}e^{-m_{\rm out}(r-R)}
\ee
and
\be
r^2 \partial_r \phi\vert_{r=R}=-(1+m_{\rm out}R) (\phi_{\rm in}-\phi_{\rm out}) {R}.
\label{gra}
\ee
Notice that this result is valid for any value of $m_{\rm out} R$, i.e. irrespectively of the presence or not of the Yukawa suppression term outside the object.
We will calculate the scalar charge in the following appendix.

\section{The Force Law}

In the main text we have used the scalar charge of screened and unscreened objects. In this appendix, we will give a more rigorous treatment following the discussion in \cite{Hui:2009kc}.
First of all let us recall that in the Einstein frame the Einstein equation reads
\be
G_{\mu\nu}= 8\pi G_N (T^m_{\mu\nu}+ T^\phi_{\mu\nu})
\ee
where we have that the energy momentum of matter is
\be
T^m_{\mu\nu}= A(\phi) \rho U_\mu U_\nu
\ee
where $\rho$ is the conserved matter density and $U_\mu$ the velocity 4-vector. The scalar energy momentum is
\be
T^\phi_{\mu\nu}= \partial_\mu \phi\partial_\nu\phi -g_{\mu\nu} (\frac{(\partial \phi)^2}{2} +V).
\ee
We work in the Newton gauge for the metric
\be
ds^2=-(1+2\Phi_N) dt^2 + (1-2\Phi_N)dx^2
\ee
and we expand both the scalar field and the Newton potential as
\be
\Phi_N= \Phi_0 +\Phi_{\rm out},\ \phi= \phi_0 +\phi_{\rm out}
\ee
where ``$0$'' denotes the background quantities and ``${\rm out}$'' the fields sourced by the objects. In the vicinity of an object, the background field can be expanded to linear order
\be
\Phi_0(\vec x)= \Phi_0(0) +\partial_i \Phi_0 (\vec x) x^i, \ \phi_0(\vec x)= \phi_0(0) +\partial_i \phi_0 (\vec x) x^i.
\ee
The field outside the given object and created by the object is such that, as long as $m_{\rm in} R\gg 1$ where $m_{\rm in}$ is the mass of the scalar field inside the object and $R$ its typical size
\be
\phi= \phi_{\rm out} + ({\phi_{\rm in}-\phi_{\rm out}}) \frac{R}{r}e^{-m_{\rm out}(r-R)}.
\ee
We also assume that the object creates the Newtonian potential
\be
\Phi_{\rm out}(r) = -\frac{G_N M}{r}
\ee
where $M$ is the mass of the object. We assume that matter is responsible for the Newtonian potential, and that the scalar field energy scale is negligible compared to matter inside the object
and very small outside.

The Einstein equation can be rewritten as
\be
G^{(1)}_{\mu\nu}= 8\pi G_N (T^m_{\mu\nu}+ T^\phi_{\mu\nu}+ + t_{\mu\nu})\
\ee
where $G^{(1)}_{\mu\nu}$ is linear in the Newton potential and the pseudo-tensor is given by
\be
t_{\mu\nu}=-\frac{1}{8\pi G_N} G^{(2)}_{\mu\nu}
\ee
where $G^{(2)}_{\mu\nu}$ contains all the higher order terms in the Newton potential. This corresponds to the gravitational pseudo energy momentum tensor.

We can now identify the expression for the mass $M$ of the object which is given by
\be
M= -\int_V d^3x \tilde T_0^0.
\ee
where we  draw a sphere of volume $V$ encircling the object and
\be
\tilde T_{\mu\nu}=T^m_{\mu\nu}+ T^\phi_{\mu\nu} + t_{\mu\nu}.
\ee
Neglecting the scalar contribution to the energy density, the mass is given by the integral over the object
\be
M= \int_V A(\phi) \rho d^3x
\ee
which is constant as long as the scalar field is time-independent and we can neglect the radiation by gravitational waves.

The momentum of the object is simply given by
\be
P_i= \int_V  d^3 x \tilde T_i^0.
\ee
The non-covariant Bianchi identity implies that $\partial^\mu \tilde T_{\mu\nu}=0$ and therefore we get
\be
\dot P_i= - \int_{\partial V} dS_j \tilde T^j_i
\ee
where the surface integral is on the surface of the outer sphere. There the matter energy momentum tensor is negligible, and similarly for the contribution from the scalar field energy density. Only two terms have a relevant flux: the scalar and gravitational ones.
The gravitational flux has been computed in \cite{Hui:2009kc} and yields
\be
\int_{\partial V} dS_j t^j_i=\frac{r^2}{G_N}\frac{\partial \Phi_{\rm out}}{\partial r}\partial_i \Phi_0= M \partial_i \Phi_0
\ee
which gives a contribution equal to the GR prediction.
The new contribution from the scalar field is simply dominated by the large gradient of the scalar field $\phi_{\rm out}(r)$ compared to the scales over which the background quantities vary
\be
-\int _{\partial V} dS_j T^{\phi j}_i= -4\pi r^2 \frac{\partial \phi_{\rm out}}{\partial r}\vert_{r=R} \partial_i \phi_0
=4\pi (1+m_{\rm out}R)(\phi_{\rm in} -\phi_{\rm out})R  \partial_i \phi_0 \,.
\ee
using (\ref{gra}) and the flux is evaluated at the outer surface of the object.
As a result we obtain that
\be
\dot P_i= -M \partial_i \Phi_0-4\pi(1+m_{\rm out}R)(\phi_{\rm out} -\phi_{\rm in})R  \partial_i \phi_0.
\ee
Now the centre of mass coordinates
\be
M X^i= -\int_V d^3 x x^i t^0_0
\ee
is such that $P^i= M \dot X^i$ and therefore
\be
\ddot X^i= -\partial^i \Phi_0-(1+m_{\rm out}R)\frac{(\phi_{\rm out} -\phi_{\rm in})}{2m_{\rm Pl}\Phi_N(R)}  \frac{\partial^i \phi_0}{m_{\rm Pl}}
\ee
where $\Phi_N (R)= \frac{G_N M}{R}$. We can immediately identify the charge of a given object
\be
\beta_{\rm object}=(1+m_{\rm out}R)\frac{(\phi_{\rm out} -\phi_{\rm in})}{2m_{\rm Pl}\Phi_N(R)}
\ee
such that
\be
\ddot X^i= -\partial^i \Phi_0 -\beta_{\rm object} \frac{\partial^i \phi_0}{m_{\rm Pl}}
\ee
which is exactly what we used in the main text.
From this we can immediately deduce that the field generated by an extended object is given by
\be
\partial^i\phi_{\rm out}= 2 \beta_{\rm object} {m_{\rm Pl}} \partial^i \Phi_{\rm out}.
\ee
This implies that when the external fields $\Phi_0$ and $\phi_0$ are due to another extended object, we find that the motion of the object $A$ is due to the total potential $(1+2\beta_A \beta_B)\Phi_B$
where $\Phi_B$ is the Newtonian potential due to a second object $B$ and such that we have
\be
\ddot X^i_A= - (1+2\beta_A \beta_B) \partial^i \Phi_B
\ee
which is also the result that we have used in the main text.

\section{The E\"otwash bound}

The E\"otwash experiment has been analysed numerically in \cite{Upadhye:2012qu,Upadhye:2012rc} and analytically in \cite{Brax:2014zta}. We follow the latter in this appendix.
The search for the presence of new interactions by the Eotwash experiment \cite{Adelberger:2003zx} involves two plates separated by a distance $d$ in which holes of radii $r_h$ have been drilled regularly on a circle. The two plates rotate with respect to each other. The gravitational and scalar interactions induce a torque on the plates which depends on the potential energy of the configuration. The potential energy is obtained by calculating the amount of work required to approach one plate from infinity \cite{Brax:2008hh,Upadhye:2012qu}. Defining by $A(\theta)$ the surface area of the two plates which face each other (this is not the whole surface area because of the presence of the holes), a good approximation to the torque, expressed as the derivative of the potential energy of the configuration with respect to the rotation angle $\theta$, is  given by
\be
T \sim a_\theta \int_d^{d_{\rm max}} dx (\frac{\Delta F_{\phi}}{A}(x))
\label{tor}
\ee
where $a_\theta=\frac{dA}{d\theta}$ depends on the experiment. The pressure $\frac{\Delta F_{\phi}}{A}(x)$ is the Casimir pressure due to the scalar field between the two plates separated by a distance $x$.
When the Casimir pressure due to the scalar field decreases fast enough with $x$, the upper bound $d_{\rm max}$ can be taken to be infinite. When this is not the case, the upper bound is the maximal distance below which the scalar force is not suppressed by the Yukawa fall-off.

 For the 2006 Eot-wash experiment \cite{Kapner:2006si}, we consider the bound obtained
 for a separation between the plates of $d \lesssim 1\ {\rm mm}$
\be
\vert T \vert \le a_\theta \Lambda_T^3
\ee
where $\Lambda_T= 0.35 \Lambda$ \cite{Brax:2008hh}. The pressure between the two plates is low $10^{-6}$ T corresponding to a redshift of  $a_b\sim 1.4\ 10^{-6}$.
We must also modify the expression of the torque (\ref{tor}) in order to take into account the effects of a thin electrostatic shielding sheet of width $d_s=10\mu {\rm m}$ between the plates. This reduces the observed torque which becomes
\be
T_{obs}=e^{-m_s d_s} T_\theta
\ee
where $m_s$ is the mass of the scalar field in the shield.
When the mass in dense media is very large, this imposes a strong reduction of the signal.

The scalar Casimir pressure has been calculated \cite{Brax:2014zta}
\be
\frac{\Delta F_{\phi}}{A}= V_{eff}(\phi_b)-V_{eff}(\phi_d)
\label{cas}
\ee
corresponding to the difference between the effective potential in vacuum $\phi_b$ compared to the value it takes in between the plates $\phi_d$ (at the mid-point between the plates). As long as the density between the plates is lower than the density in the plates, the scalar field form bubbles reaching a value $\phi_d\ne \phi_b$.
This can also be expressed using the tomographic mapping as
\be
\frac{\Delta F_{\phi}}{A}=-27 \Omega_{m0}^2 {H_0^4 m_{\rm Pl}^2}\int_{a_d}^{a_b} da \frac{\beta^2(a)}{ a^7 m^2(a)}(1-\frac{a^3 }{a_b^3})
\ee
where $\rho_b=\frac{\rho_0}{a_b^3}$ and $\phi_d=\phi(a_d)$. Hence the scalar field adds an extra attracting pressure between the plates as the integrand is always positive.
The value of $\phi_d$ depends on the masses $m_{\rm plate}$ and $m_b$ in the plates and in the vacuum. When $m_{\rm plates} d\gtrsim 1$, the field has a non trivial profile between the plates, i.e. there is bubble of scalar field,  and the scalar Casimir pressure does not vanish. When $m_{\rm plate} d \lesssim 1$, the field is constant between the plates and $\phi_d=\phi_{\rm plate}$ implying a constant scalar Casimir pressure. Finally when the plates are not screened and $m_{\rm plate} D_{\rm plate}\lesssim 1$ where $D_{\rm plate}$ is the width of the plates, we have $\phi_d= \phi_b$ and no Casimir pressure is present.

The case of chameleons and symmetrons can be found explicitly treated in \cite{Brax:2014zta}. Here we repeat our results for the symmetrons as we have used them in figure (\ref{syy}).
In the symmetron case, we find a constant pressure for close enough plates and no pressure when they are far apart. Indeed as long as $m_b d\le {\sqrt 2}{\pi}$, we have that
\be
\phi_d=0
\ee
and the Casimir pressure is given by a constant
\be
\frac{\Delta F_{\phi}}{A}=-\frac{\mu^4}{4\lambda}
\ee
which is the height of the symmetron potential. We have assumed that the vacuum is perfect between the plates. When $d\gtrsim d_c=\frac{\sqrt 2 \pi}{ m_b}$, the interaction between the plates is Yukawa suppressed implying that
we can approximate it to be vanishing.
Hence the torque is given by
\be
T_\theta=- a_\theta \frac{\mu^4 ( d_c-d)}{4\lambda}
\ee
which depends on $\mu$ and $\lambda$.

The mass of the scalar field in the shield of density $\rho_s$ is given by
\be
m_s^2= \frac{\rho_s}{M_\star^2} -\mu^2 .
\ee
As a result the
 Eot-wash bound can be expressed as
\be
M_\star\le \frac{\sqrt{\rho_s} d_s}{(\ln^2(\frac{\mu^4(d_c-d)}{4\lambda \Lambda_T^3})+\mu^2 d_s^2)^{1/2}}
\ee
as long as $4\lambda \Lambda_T^3 \le \pi \mu^4 (d_c -d)$, i.e. the upper bound on $\lambda$ in figure 3.

The torque calculation that we have presented  applies only when $a_\star\ge a_{\rm plate}$ where
\be
a_\star=(\frac{\rho_0}{\mu^2 M_\star^2})^{1/3}.
\ee
For larger values of $M_\star$ we have $a_\star \lesssim  a_{\rm plate}$ and therefore the symmetron is nearly in its vacuum phase in the plates and in the vacuum. This leads to hardly any torque between the plates.

\end{document}